# Fast and Scalable Distributed Deep Convolutional Autoencoder for fMRI Big Data Analytics

Milad Makkie
Computer Science and Bioimaging Center, The University of Georgia, Athens, GA, USA, milad@uga.edu

Heng Huang
School of Automation, Northwestern Polytechnical University, Xi'an, 710072, China, huangheng2014@gmail.com

Yu Zhao
Computer Science and Bioimaging Center, The University of Georgia, Athens, GA, USA, zzhaoyu@uga.edu

Athanasios V. Vasilakos
Lab of Networks and Cybersecurity, Innopolis University, Tatarstan, Russia, t.vasilakos@innopolis.ru

Tianming Liu
Computer Science and Bioimaging Center, The University of Georgia, Athens, GA, USA, , tliu@uga.edu

## ABSTRACT

In recent years, analyzing task-based fMRI (tfMRI) data has become an essential tool for understanding brain function and networks. However, due to the sheer size of tfMRI data, its intrinsic complex structure, and lack of ground truth of underlying neural activities, modeling tfMRI data is hard and challenging. Previously proposed data modeling methods including Independent Component Analysis (ICA) and Sparse Dictionary Learning only provided shallow models based on blind source separation under the strong assumption that original fMRI signals could be linearly decomposed into time series components with corresponding spatial maps. Given the Convolutional Neural Network (CNN) successes in learning hierarchical abstractions from low-level data such as tfMRI time series, in this work we propose a novel scalable distributed deep CNN autoencoder model and apply it for fMRI big data analysis. This model aims to both learn the complex hierarchical structures of the tfMRI big data and to leverage the processing power of multiple GPUs in a distributed fashion. To deploy such a model, we have created an enhanced processing pipeline on the top of Apache Spark and Tensorflow, leveraging from a large cluster of GPU nodes over cloud. Experimental results from applying the model on the Human Connectome Project (HCP) data show that the proposed model is efficient and scalable toward tfMRI big data modeling and analytics, thus enabling data-driven extraction of hierarchical neuroscientific information from massive fMRI big data.

## 1: INTRODUCTION

The sheer complexity of the brain has forced the neuroscience community and particularly the neuroimaging experts to transit from the smaller brain datasets to much larger hard-to-handle ones. The cutting-edge technologies in the biomedical imaging field, as well as the new techniques in digitizing, all lead to collect further information from the structural organization and functional neuron activities in the brain through rich imaging modalities like fMRI [1]. Projects such as Human Connectome Project (HCP) [2], 1000 Functional Connectomes [3] and OpenfMRI [4] are the perfect examples of such large neuroimaging datasets. The primary goal of these efforts is to gain a better understanding of the human brain and to diagnose the neurological and psychiatric disorders. Among various neuroimaging methods, task-based functional magnetic resonance imaging, tfMRI, has been widely used to assess functional activity patterns and cognitive behavior of human brain [5, 6, 7, 8]. However, the main challenges are to obtain meaningful patterns from the intrinsic complex structure of tfMRI and also lack of clear insight into the underlying neural activities. Given the hierarchical structure of functional networks in human brain, the previously data-driven methods such as Independent component analysis (ICA) [9] and sparse coding for Dictionary Learning[10] as well as model-driven approaches such as General Linear Model (GLM) [11] have been demonstrated to disregard some of the information contained in the rich tfMRI data [12, 13]. Thus, these shallow machine learning models are not capable of fully understanding the deep hierarchical structures of functional networks in human brain [12, 13]. Consequently, there is an urgent call for more efficient and scalable data analytics and knowledge discovery methods to crack the underlying brain activities.

Recently, new data-driven computational intensive neural network approaches such as deep learning have gained increasing interest among researchers, due to their efficiency of extracting meaningful hierarchical features from the low-level raw data. Particularly, Convolutional Neural Network (CNN) is among the top deep learning methods in the scientific community [14, 15, 16, 17, 18, 19, 20], especially in classifying and learning image data [21].

In the context of high dimensional data such as fMRI, however, the large size of training examples (dozens of millions of time series each with hundreds of time points) and the sheer size of model parameters can drastically impact the computational cost and accuracy of learning the fMRI signals. Furthermore, most of the current neural network methods for fMRI analysis are only implemented for local application without any parallelization scheme [28, 29, 30, 31, 32]. As indicated by an extensive battery of literature [22, 23, 24, 25], many of scaling deep learning applications by using large-scale clusters of GPUs can solve the computational bottleneck for efficient and effective knowledge discovery from fMRI big data.

Following the previous successes in using distributed GPU processing for scaling neural network model, in this work we aim to design a fast and scalable distributed framework and to implement a deep convolutional model, dubbed distributed Deep Convolutional Autoencoder (dist-DCA) to leverage the power of distributed optimization, distributed data partitioning, and multiple GPU processing. The distributed optimizer is based on an asynchronized Stochastic Gradient Descent (SGD) method [22]. In this model, we have used multiple replicas of a single model to optimize parameters, which lead to reducing the training time significantly. For data parallelization, we utilized Apache Spark [27] and Hadoop Distributed File System (HDFS). Considering the computationally intensive operations in tuning the parameter, Spark acts as a fast extract transfer load layer to optimize the data partitioning for the underlying Hadoop ecosystem. This is being accomplished via constructing the Resilient Distributed Dataset, RDD, which provides a functional interface to partitioned data across the cluster. Our major contributions of this work can be summarized as follows.

1) We implement a distributed deep learning framework using TensorFlow on Spark to take advantage of the power of distributed GPUs cluster.
2) We propose a distributed deep convolutional autoencoder model to gain meaningful neuroscience insight from the massive amount of tfMRI big data.
3) We validate our proposed dist-DCA model using a novel high-level sparse dictionary learning method.

Compared to the existing distributed deep learning frameworks such as dist-keras [40], elephas [47] and dl4j [48], our proposed framework has a few critical advantages: 1) The migration from a standalone code to a distributed version can be done with only a few lines of change. 2) Despite of the previous framework, our framework works efficiently with HDFS, allowing Spark to push datasets. 3) Integrating the model with the current pipeline is easy as Spark is in charge of parallelizing the data. 4) The framework is easy to deploy and scale over the cloud or in-house clusters. We have created an Amazon Machine Image (AMI), which in combination with a spark-ec2 script, can easily scale up the cluster.

The rest of this paper describes our dist-DCA model and architecture in detail. In section 2, we briefly introduce the primary components in which dist-DCA is implemented. We also review related works in this domain. We then thoroughly describe our deep convolutional model in section 3. Section 4 is dedicated to data parallelism and distributed optimization. Section 5 describes our scalable experiments in large GPU clusters where we explain how our model can be easily distributed among dozens of GPU nodes to reduce computational time efficiently.

## 2: PRELIMINARY AND RELATED WORKS

Recent advances in building affordable high-performance GPUs with thousands of cores were one of the critical factors in advancing large-scale deep learning models. This breakthrough has also encouraged the scientific community to utilize GPUs more often, as CPU's capacity does not seem to grow in proportion to the rate of increasing demand. However, the limited memory capacity of typical GPUs on the market (usually 8 gigabytes) has become a bottleneck in feeding extensive datasets as far as the training speed is concerned. Therefore, two common approaches of data parallelism and model parallelism are of the researchers' interest.

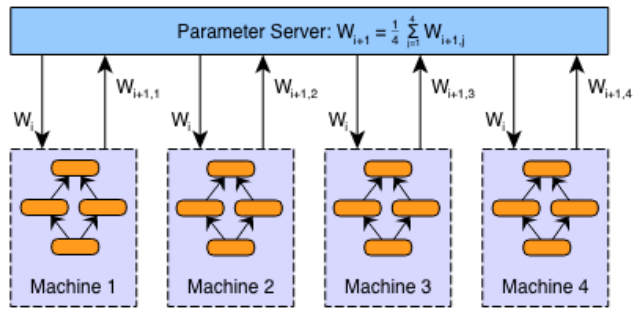

**Fig. 1: An asynchronous data parallelism model using Asynchronous SGD**

In model parallelism, different portions of a model computation are done on different computational devices simultaneously for the same batch of examples while sharing the parameters among devices as a single model. This approach, however, is efficient for very large models as splitting a neural network model needs to be done in a case-by-case manner and is very time-consuming. Data parallelism, on the other hand, seems more straightforward for general implementation and can be easily scaled to larger cluster sizes. Fig. 1 demonstrates a data parallelism paradigm. We will discuss our dist-DCA data parallelism scheme in more depths in section 4.

Our main motivation behind this work is to implement a scalable, asynchronous data parallelism model leveraging TensorFlow on Spark [34] to efficiently learn meaningful hierarchical abstraction of massive size of fMRI data.

## 2.1 TensorFlow

TensorFlow [46] is a mathematical software and an open-source software library for Machine Intelligence, developed since 2011, by Google Brain Team and initially aimed to machine learning research and deep neural networks. TensorFlow is a numerical computation library using data flow graphs that enables machine learning experts to do more data-intensive computing, e.g., it contains some robust implementations of conventional deep learning algorithms. It offers a very flexible architecture that enables deploying computation to one or more CPUs or GPUs in a standalone, parallel or distributed fashion. We selected TensorFlow in our work as it efficiently supports distributed and parallel GPU processing and it supports Keras. However, having an easy to scale framework is required for running TensorFlow applications when the model and data become large. So, a queuing framework to both seamlessly feed data into the cluster nodes and to schedule and manage tasks efficiently is needed. Pipelining pre-processing, training and inferences steps is a known challenge yet to be addressed by the TensorFlow ecosystem.

## 2.2 Spark

Since 2009, the Spark framework [27] was developed at the University of Berkeley AMPlab and currently is being maintained by Databricks. This framework addresses deficiencies of MapReduce by introducing resilient distributed datasets (RDD) abstract where the operations are performed in the memory. Spark compiles the action lineages of operations into efficient tasks, which are executed on the Spark engine. Spark offers a functional programming API to manipulate Resilient Distributed Datasets (RDDs). RDDs represent a collection of items distributed across many computing nodes that can be manipulated in parallel. Spark Core is a computational engine responsible for scheduling, distributing and monitoring applications. It consists of many computational tasks across executor node(s) on a computation node/cluster. Spark's scheduler will execute the duties across the whole cluster. Spark minimizes the repetition of data loading by caching data in memory, which is crucial in complex processes.

Spark uses Hadoop filesystem as a core distributed file system (HDFS). Apache Spark is one of the most active Apache projects on GitHub.

In this work, we used a combination of TensorFlow and Spark [46] to leverage the data parallelism and scheduling of Spark, thus enabling direct tensor communication among TensorFlow executors and parameter server(s). Process-to-process direct communication enables TensorFlow program to scale effortlessly. In section 4, we will describe such communication in more details.

## 2.3 Previous Works

In the past few years, there have been multiple studies in adopting neural network methods to model fMRI data and its associated applications. For instance, Chen et al. [28] used convolutional autoencoder in fMRI data aggregation; Plis et al. [29] used deep belief network (DBN) to learn physiologically important representations from fMRI data; Suk et al. [30] combined the Deep Auto-Encoder with Hidden Markov Model to investigate the functional connectivity in resting-state fMRI; Huang et al. [31] used the restricted Boltzmann machine to mine the latent sources in task fMRI data; Ren et al. [32] used convolutional neural networks to classify fMRI-derived functional brain networks, and Wen et al. [33] have used AlexNet to reconstruct the visual and semantic experiences using fMRI data. In the context of applying deep learning applications to fMRI data, however, most works have focused on the classification problem by using a single computation node. Our focus in this paper is the provision of an unsupervised distributed CNN encoder that effectively models the tfMRI big data. This enables us to learn hierarchical feature abstraction while lowering the spatial and temporal noises contained in fMRI data and ensures us to efficiently reduce model training and inferences time by easily scaling cluster of GPUs.

## 3: DEEP CONVOLUTIONAL AUTOENCODER

Fig. 2. illustrates both the structure of our proposed dist-DCA model and a validation pipeline based on the online dictionary learning (ODL) algorithm. We will describe this pipeline later in section 5. A neuroinformatics platform [35] is used to preprocess the tfMRI signals. Then Keras and Tensorflow APIs are used to construct the DCA model. In section 4, we will explain how asynchronous gradient computation reduces the model's training time by communicating and updating parameters values.

A non-distributed version of DCA model is elaborated in [43]. To facilitate the understanding of the model, we recapitulate the model in the following paragraphs. The purpose of autoencoder in DCA is to first encode the input fMRI time series by mapping them into higher level feature maps and then to decode the signals by reversing the process. Throughout this process, we obtain a hierarchical abstraction of fMRI signals while denoising them. As mentioned below, we assume that the model consists of only one convolutional layer both at the encoder and decoder and later we extend it to the real model. A summary of the key model parameters is shown in Table 1.

$$z_i = f(p_i * x + b_i) \qquad (1)$$

where x is the signal input, and $p_i$ and $b_i$ are the corresponding filter and bias for the i-th feature map. f is the activation

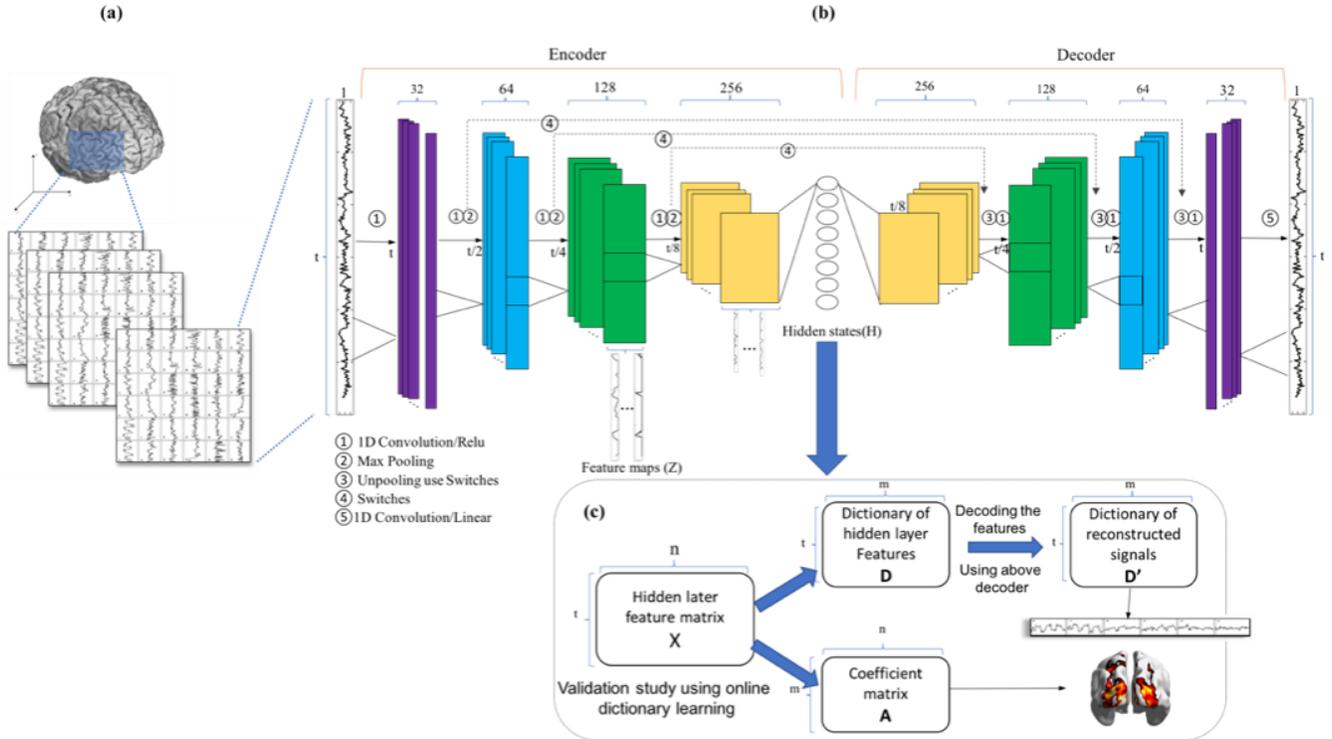

**Fig. 2:** An illustration of the dist-DCA model and the online dictionary learning validation study. (a) demonstrates the preprocessing step of the tfMRI data including signal extraction and normalization. (b) demonstrates the structure of the dist-DCA model and its components including all hidden layers and feature maps. (c) shows the validation study through which we obtain the brain activity pattern.

**TABLE 1: dist-DCA model summary**

| Feature map/filter | Layer1 | Layer2 | Layer3 | Layer4 |
|---|---|---|---|---|
| Encoder | 32/21 | 64/9 | 128/9 | 256/9 |
| Decoder | 128/9 | 64/9 | 32/9 | 1/21 |
| Total Parameters | 6,023,549 | | | |

### 3.1 Encoder:

The Encoder takes one 1D tfMRI signal x as shown in Fig. 2.b and then by convolving the filters throughout the entire signal generates the feature map in the next layer using the equation 1.

function. In this paper, except for the output convolutional layer in the decoder layer where we use linear activation function, we use the Rectified Linear Unit (ReLU) as activation functions. The advantages of choosing ReLU in our study is first to reduce the possibility of vanishing gradient and second to represent the signal sparsely as we later use the sparse representation of the hidden layer for data validation. A fully connected layer is used at the end of the encoder to match the encoder final hidden layer feature size with the input signal and to ensure that the hidden states are learned with a full receptive field of input as we use it as the final desired output of the model as mentioned in [43].

$$H = Z \times W + C \qquad (2)$$

In the equation 2, the hidden layer states are represented by H, whereas Z, W and C are the feature maps, weight and bias of the fully connected layers, respectively.

### 3.2 Decoder:

The decoder is following a symmetric pattern and attached to the previous encoder. To reconstruct the input signal, first, the hidden states are mapped and reshaped to a reconstructed version of feature maps Z′ via fully connected layer in the decoder. In equation 3, W′ and C′ denote the weights and bias of the fully connected layer in the decoder, respectively.

$$Z' = H \times W' + C' \quad (3)$$

In the end, input signal will be reconstructed by linearly combining these feature maps, where $\hat{x}$ denotes the reconstructed signal, and $p_i'$ and $b_i'$ are the filters and biases in the decoder as shown in eqation 4.

$$\hat{x} = \sum_i p_i' * z_i' + b_i' \quad (4)$$

The same concept is extended to a model with more layers (4 layers in encoder and 4 in the decoder) by transforming the input layer into different feature map in each convolutional layer by a chain rule. To minimize the mean square error between fMRI signals and their reconstructions, we also used an L2 regularization term between feature maps in the top layer of the encoder and the bottom layer of the decoder. Doing so ensures us that the fully connected layer does not randomly shuffle the timing order when reconstructing features maps in the decoder. λ in equation 5 controls the significance of the L2 regularization term and we experimentally set it to 0.006.

$$\min \frac{1}{2}||X - \hat{X}||_2^2 + \frac{1}{2}\lambda ||Z - Z'||_2^2 \quad (5)$$

### *3.3 Max-pooling and unpooling:*

The max pooling is applied on each layer after the convolutional layer. This helps first by substantially reducing the computational cost for the upper layer and second, by gaining translation-invariance. The translation-invariance is particularly important in tfMRI due to possible time-shift phenomena while acquiring the raw signal [38, 34].

Given the invertible property of max-pooling, we utilized switches [39] in the encoder to memorize the location of the local max in each pooling regions and then we applied the location of the corresponding local max value to its original position. In validation studies (section 5) when "switches" are not available, we simply use traditional up-sampling.

In the next section, we explain how such a model is replicated among spark executor nodes.

## 4: DATA PARALLELISM AND MODEL DEPLOYMENT

In data parallel approaches, a copy of the entire model is sent to each executor, parallelizing processing of gradient descent by partitioning data into smaller subsets. A parameter server then combines the results of each subset and synchronize the model parameters between each executor after receiving gradient delta from each executor. This can be done synchronously or asynchronously. However, in homogeneous environments where nodes share the same hardware specifications and communicate via a reliable network of communication, asynchronous [22] methods outperform for two reasons [22, 26]. First, executors do not wait for others to commit before start processing the next batch of data. Second, asynchronous method is more robust to failure of nodes as if one node fails the others will still train their own data partitions and fetch new updates from parameter server.

For example, given a batch size of 100 elements, 5 replicas of the model computes the gradient for 20 elements, and then combine the gradients in a separate node, known as parameter server, and apply parameters updates synchronously, in order to behave exactly as if we were running the sequential SGD algorithm with a batch size of 100 elements.

We have implemented downpour SGD [22] in our distributed framework and have fixed the $\eta_{fetch}$ and $\eta_{push}$ of weights and gradients to one, for speeding up convergence and for ease of comparison to simple SGD. Our experiment shows that relaxing consistency requirements are remarkably effective. Downpour SGD comes from the intuition that if we view the gradient descent as a water droplet toward minimizing the error rate, then individual executors can be considered as several droplets near each other, all separately flowing down into the same valley. Moreover, we practiced a warm-up phase, wherein a single executor node starts training on its own data partition before starting other executors. This has significantly decreased the probability of diverging of each executor being trapped in its own local optima.

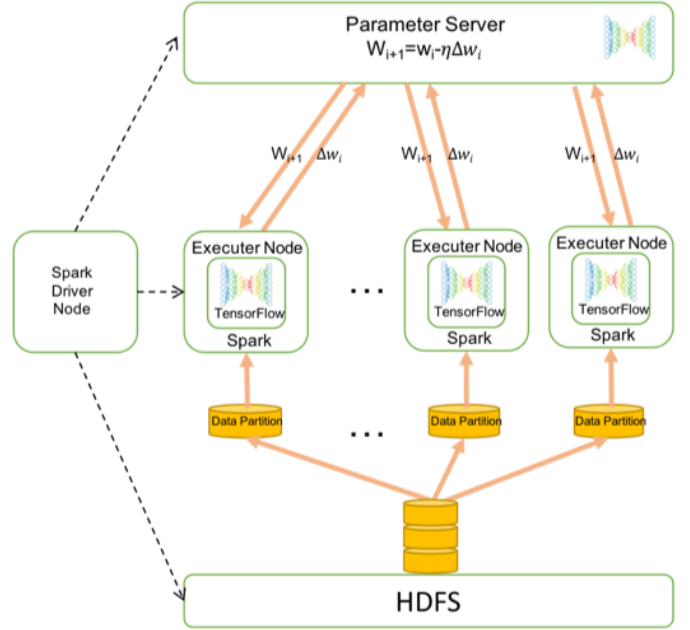

**Fig. 3: Dist-DCA. Executor nodes asynchronously fetch parameters w and push gradients to the parameter server. Spark driver is also in charge of data penalization and task scheduling.**

We also chose the Adagrad optimizer [41] to keep the learning rate update for each parameter as the model is training and to ease extending the number of executing nodes. Adagrad uses a separate adaptive learning rate for each parameter. Let $\eta_{i,K}$ be the learning rate of the i-th parameter at iteration K and $\Delta w_{i,K}$ its gradient, then in equation 6 we obtain $\eta_{i,K}$.

$$\eta_{i,K} = \gamma / \sqrt{\sum_{j=1}^{K}(\Delta wi,j)^2} \qquad (6)$$

Because these learning rates are computed only from the summed squared gradients of each parameter, Adagrad is easily implemented locally within the parameter server. $\gamma$ is the constant scaling factor for all learning rates, which is larger than the best fixed learning rate used without Adagrad. The use of Adagrad extends the maximum number of model replicas that can productively work simultaneously.

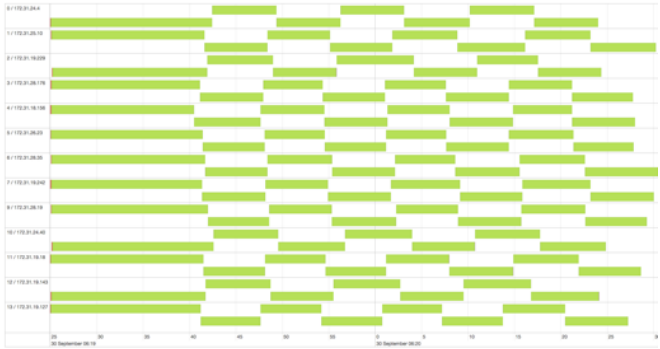

**Fig. 4: Dist-DCA data partitions. Spark driver keeps track of data partitions and executors' computational times. Here you can see the computational time of all active nodes and how the training tasks are scheduled.**

The abovementioned optimization procedures ideally address our problem in two ways. One is by empowering us to process massive fMRI data (nearly 10 millions tfMRI time series each with 284 time points as in this work). And, the other is by allowing us to train our relatively large model, consisting of more than 6 million trainable parameters, faster. As a result, our proposed dist-DCA benefits from asynchronous data parallelism through two main components of distributed data partitioning and distributed parameter optimization as it is shown in the Fig. 3. We use Hadoop as our main distributed file system and Spark for tasks scheduling and data partitioning. Each Spark executor acts as a wrapper of Tensorflow application where one node handles the parameter synchronization and the rest run the Tensorflow application independently just as one single node setup. Each executor commits its gradient delta to parameter server after each processing batch elements and receives the latest parameter from the server. Meanwhile, Spark core efficiently feeds each of the executors through HDFS by partitioning the data based on the number of epochs and dataset size. The Fig. 4. shows Spark data partitioning among a cluster of 16 nodes consisting of one driver, two parameter servers, and 13 executors. Spark driver is responsible for handling tasks and for replicating TensorFlow model across a cluster. For each stage and each partition, tasks are created and sent to the executors. If the stage ends with a shuffle, the tasks created will be shuffle-map tasks. After all tasks of a particular stage are completed, the driver creates tasks for the next stage and sends them to the executors, and so on. This repeats until the last stage, where the results return to the driver. With the asynchronized implementation, we ensure that both the model replicas and data partitions are run independently, thus reducing the delays induced by the loaded executors.

## 5: EXPERIMENTS

We evaluated the TensorFlow on Spark performance and scalability by our novel dist-DCA model using Amazon Elastic Cloud Computing (EC2). We trained this model on 9,658,464 fMRI time series of 48 human subjects and evaluated on 4,024,360 time series of 20 separate subjects.

### 5.1 Experiments Setup

*5.1.1 Dataset.* We use the Human Connectome Project (HCP) Q1 release dataset [36] containing 68 healthy subjects' tfMRI data. The HCP dataset is advantageous in its high temporal and spatial resolution (TR=0.72s, varied temporal length from 176 to 1200 volumes; 2mm isotropic voxels, to the total of over 201,218 voxels' signals per subject each with the length of 284 time points), which enables more detailed characterization of the brain's functional behaviour. We use motor task fMRI data in this study, composed of six most basic motor tasks including visual cues (event 1), tapping left (right) fingers (event 2, 3), squeezing left toes (event 4, 5) and moving tongue (event 6). We divided the Motor task Q1 subjects into two separate subsets of 48 training and 20 validating subjects. For running the dist-DCA model, the preparation steps include fMRI signal pre-processing (gradient distortion correction, motion correction, bias field reduction, and high pass filtering) [37], all implemented using FSL FEAT. Furthermore, we recruited our integrated neuroinformatic platform, HELPNI [35], to facilitate the pre-processing and to integrate different steps of data acquisition using its powerful pipelining ability.

*5.1.2 Cloud Platform.* The dist-DCA model is deployed on Amazon Web Service Elastic Cloud Computing, AWS EC2. EC2 clusters are highly scalable, as the number of executor nodes could be adjusted effortlessly within the cluster. The pre-processed and converted fMRI data was stored in the cloud through Amazon S3 and accessible by the EC2 clusters. This enables us to pull data to newly initialized instances easily. We used customized scripts along with an AMI containing a preconfigured instance to scale our cluster according to desire. Each cluster's node contains Apache Spark version 2.2.0, Hadoop version 2.6.0, TensorFlow 1.3, Keras 2.08 and python 2.7. To benchmark the scalability and robustness of our proposed framework, we used a variety of node hardware configurations with a different number of node per experiment as summarized in Table 2. The configuration of nodes are as follows. G3 nodes are equipped with High-Frequency Intel Xeon E5-2686 v4 (Broadwell) processors, NVIDIA Tesla M60

GPU, with 2048 parallel processing cores and 8 gigabytes of video memory per GPU with 25 Gbps of aggregate network bandwidth within the cluster. G2 nodes come with Intel Sandy Bridge processors, NVIDIA Kg20 Grid GPU with 1536 CUDA cores and 4 gigabytes of memory per GPU.

**TABLE 2: Cloud clusters' configuration, each line represents a separate experiment setup.**

| No of Spark/TF Executors | vCPU Cores/node | Used GPU Memory per node (GB) | Memory per node (GB) | EC2 Node Type |
|---|---|---|---|---|
| 1 | 16 | 8 | 122 | G3-4x |
| 2 | 16 | 8 | 122 | G3-4x |
| 4 | 4 | 12 | 61 | P2-x |
| 4 | 8 | 4 | 15 | G2-2x |
| 4 | 16 | 8 | 122 | G3-4x |
| 4 | 32 | 4 | 60 | G2-8x |
| 8 | 16 | 8 | 122 | G3-4x |
| 13 | 16 | 8 | 122 | G3-4x |

### 5.2 Performance

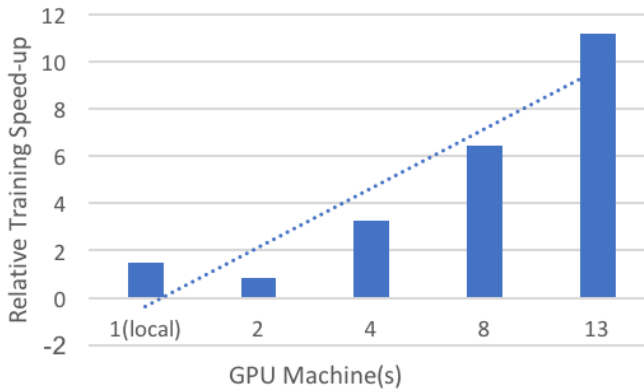

**Fig. 5: Training speed-up versus the numbers of local/spark executors.**

We aimed to investigate the performance of our framework with respect to the mean processing time of a single mini-batch (1 fMRI signal) for Downpour SGD with Adagard training as a function of the number of nodes used in a single model instance. To do so, we deployed four clusters of G3 instances with 4, 6, 8 and 16 nodes (correspond to 2, 4, 8 and 13 TensorFlow executors respectively as shown above). Given the broadband network communications, except for the 16-node cluster with two parameter servers, we only dedicated one parameter server along with one spark driver. Moreover, to evaluate the effect of network traffic on training speed, we ran a non-distributed version (called DCA) of the model on a single node with the same configuration. In all the experiments, we trained our models (dist-DCA and standalone DCA) on 9,658,464 time series of HCP Q1 data for 1600 batches and 6036 steps per epoch.

Fig. 5. demonstrates the speed of various implementations including our standalone and distributed ones on GPU nodes. Since the standalone DCA has no data-parallelism and no network overhead, it obviously outperforms the two-node cluster. However, clusters with the higher number of executor nodes easily exceed regarding computation time. For example, the cluster with 13 executors outperforms the standalone model with almost seven times. It can be observed that training speed linearly grows as the number of executor nodes increase. However, we expect that the performance drops if we increase the number of executor nodes to more than 13. This happens as network overhead starts to rule over our dist-DCA model performance and as executor nodes will have fewer tasks to do while waiting to fetch new parameters.

### 5.3 Scalability

To further demonstrates the scalability of our implemented distributed framework, we measured the training time of dist-DCA on the previously discussed dataset. We trained our model in 4 different cluster settings with 2, 4, 8 and 13 G3 executor nodes with a total of 2, 4, 8 and 13 GPUs respectively. Please note that for the sake of comparison, in all experiments, only one GPU per node was used. Our goal is to obtain minimum loss in the minimum amount of training time. Fig. 6. illustrates that the training time is reduced significantly by almost 51 hours in a four-executor with 64 CPU cores (4 GPUs) compared to a two-executor cluster with 32 CPU cores (2 GPUs). However, this increased rate does not hold from the four-node to the eight-node cluster with 128 cores of CPU (8 GPUS). We believe that this is due to network communication overhead and previously discussed warm-up phase. As explained in section 4, the TensorFlow application (here dist-DCA model) is wrapped inside a spark executor at each node. Executors independently start to train the model by pushing gradients and fetching the new parameters from the parameter server at each stage. These recurring network communications can cause the larger clusters to not linearly scale-up as opposed to the ones with fewer nodes. We can conclude that network can always be a bottleneck in larger clusters.

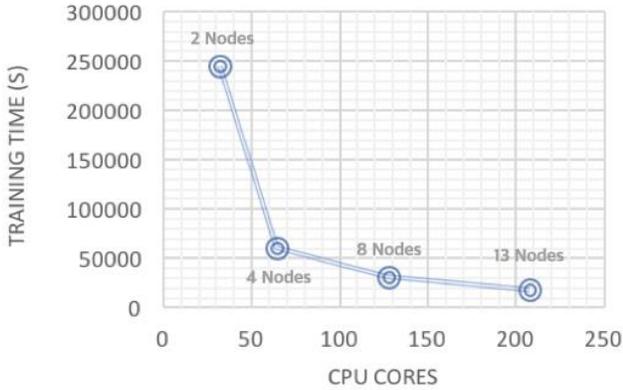

**Fig. 6: Training time of dist-DCA based on the number of CPU cores on different cluster setups.**

We also performed another experiment solely to evaluate the effect of CPU cores on our proposed framework performance. To do so, we launched 4 clusters each with 4 nodes to train dit-DCA model over our HCP data. In each cluster setup, we used the same environmental setup and one GPU card per node (total of 4 GPUs). Clusters utilized 4, 8, 16 and 32 CPU cores per node respectively. Demonstrated results in Fig. 7 suggests that increasing only the number of CPUs would not benefit the training speed significantly. A simple comparison of the results in Fig. 5 with the Fig. 7 shows that while increasing the number of GPUs in a distributed setup reduces training time significantly; such a conclusion cannot be drawn as opposed to increasing CPU cores.

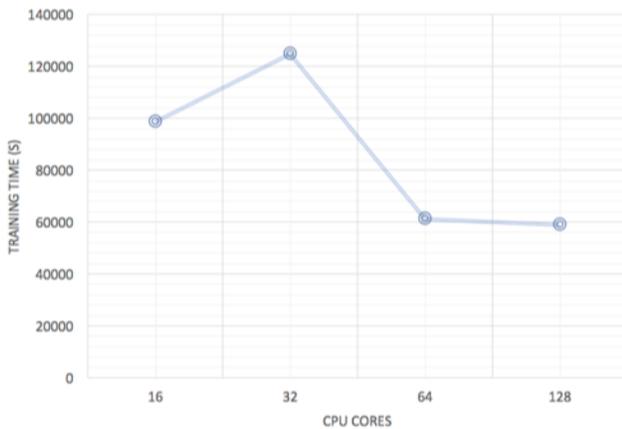

**Fig. 7: Training time of dist-DCA based on the number of CPU cores in clusters with the same number of nodes.**

### 5.4 Learned Features Validation and Visualization

To validate the learned features of our proposed model, we have performed a validation study on the hidden layer features of the encoder. An illustration of the computational procedure of this validation study is shown in the Fig. 2c. The rationale behind this is to compare the detected task-related patterns of brain activity through a sparse dictionary learning method in two setups. One by feeding the high-level features of the hidden layer (setup 1) contained by dist-DCA and the other with the raw tfMRI signals (setup 2). Sparse dictionary learning as an unsupervised learning algorithm aims at finding a sparse representation of input data in the form of a linear combination of basic elements, known as dictionaries along with their corresponding coefficients. [10, 44, 45]. This goal is achieved by aggregating fMRI signals into an over-complete dictionary matrix and a corresponding coefficient matrix through an effective online dictionary learning algorithm [42]. The time series of each over completed dictionary represents the temporal activity of a brain network, and its corresponding reference weight vector stands for the spatial map of every network. This method is recognized as an efficient method for inferring a comprehensive collection of concurrent functional networks in the human brain [10]. The spatial and temporal pattern of a sample network decomposed results is demonstrated in Fig. 8.

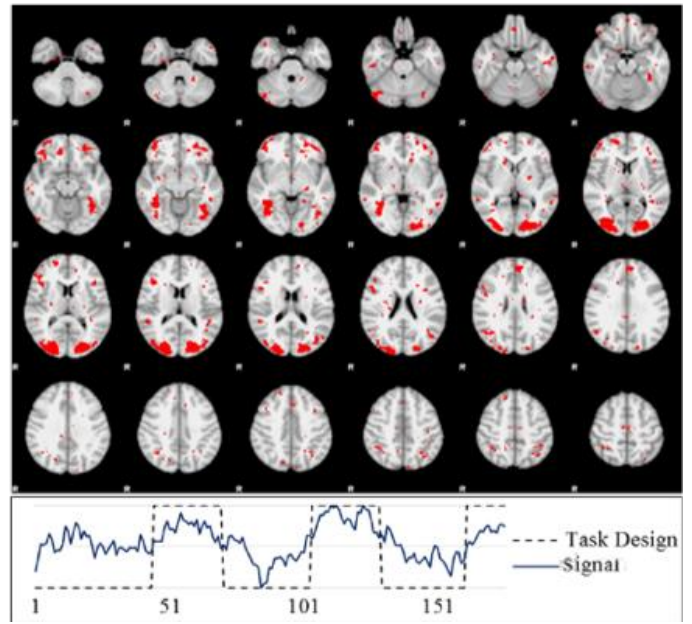

**Fig. 8: Top: spatial pattern visualized on cortical volumetric space of one decomposed network. Bottom: visualization of its temporal pattern.**

To draw a fair comparison, we have used the same parameters in both runs. We adopted the parameter-tuning approach that Lv et al. suggested [10]. Both setups learned 400 dictionaries with 0.7 sparsity regularizer [42] to achieve the best performance of the brain network inference. After training, the high-level features of setup one are decomposed as high-level dictionaries and corresponding spatial distributions. Then we use the decoder to project these high-level dictionaries (time series patterns) back to the signal space. The detected patterns are visualized in Fig. 9. As shown on the right side of the figure, although the dictionary learning analysis in both setups has detected all the six motor task patterns, these patterns are mixed with a large number of noises in setup 2, and as a result, the correlation values with task design pattern are quite small. On the other hand, the setup 1 contained much fewer noises in both of the time series patterns and spatial maps. Consequently, we can conclude that our proposed model filters noises in each layer and preserves the useful information of the brain activities. For the sake of simplicity and page limitation, we do not explain the theoretical brain model analysis and reconstruction error analysis. Further details of this comparison can be found at Huang et al. work [43].

Furthermore, we visualized the filters in each layer. Fig. 10 shows all 32 filters in the first layer of the encoder. The first layer filters summarized the most common sub-shapes of tfMRI time. For example, sinuous and bowl patterns of fMRI are shown with arrows at Fig. 10.

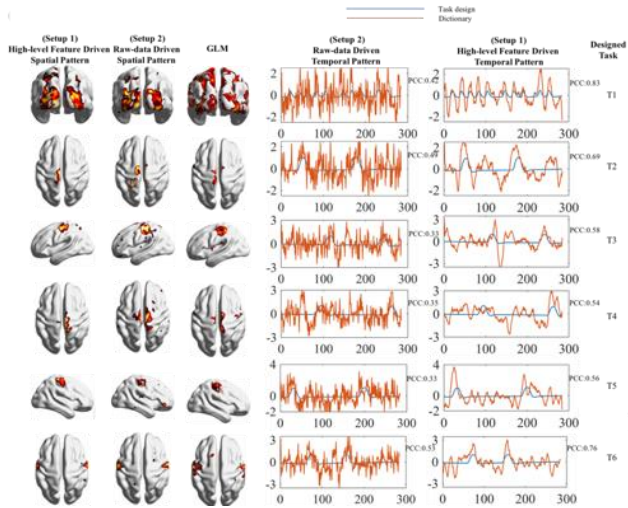

**Fig. 9: Validation study of the dist-DCA. Comparing the temporal and spatial patterns of 6 motor tasks driven from high-level features and raw-data. GLM is for reference. Pearson correlation of the designed tasks with learned dictionary atoms is shown as PCC value**

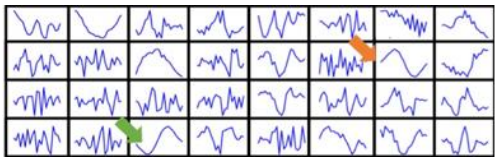

**Fig. 10: All 32 filters in the first layer of encoder. Arrows show the most common pattern of tfMRI time series.**

## 6: CONCLUSION

Providing an effective model to represent the large scale tfMRI data to break down the intrinsic complex structure of tfMRI signals has been highly demanded yet challenging. A novel deep learning model along with distributed computing are the keys to transforming our understanding of some of the most complicated brain signals [43]. In this work we presented a novel scalable distributed deep convolutional autoencoder that hierarchically models large-scale tfMRI time series data while gaining a higher level abstraction of the tfMRI signal. We used Apache Spark and TensorFlow as the computational engines to parallelize millions of fMRI time series and to train our model over large cluster of GPUs. Our experiment results showed that such a model can effectively scale-up to dozens of computation nodes, processing extensive dataset over hundreds of computational cores. The significance of network overhead, however, can severely impact the training time. Furthermore, our results showed that the high-level features are superior in task-related regions detection. The proposed autoencoder was also able to denoise the tfMRI signal as the learned dictionary atoms by our novel high-level sparse dictionary learning suggests. In general, our work contributes a novel deep convolution autoencoder framework for fMRI data modelling with significant application potentials in cognitive and clinical neuroscience in the future.

In our future work, we plan to perform further tests to implement a parallel version of our model to use the computational power of multi-GPU on a multi-node distributed setting to maximize the performance. We also plan to use the 1200+ available subjects of all HCP releases including acquisitions of different types of tasks to identify brain areas in a wide range of neural systems (such as Relation, Working Memory, Language, Social Interaction, Motor, etc.). This will benefit from our proposed distributed model, enabling data-driven hierarchical neuroscientific discovery from massive fMRI big data in the future.

## ACKNOWLEDGMENT

This work was supported by National Institutes of Health (DA033393, AG042599) and National Science Foundation (IIS-1149260, CBET-1302089, and BCS-1439051). Tianming Liu is the corresponding author of this work; phone: (706) 542-3478; E-mail: tianming.liu@gmail.com.